\documentclass[reprint,amsmath,amssymb,aps]{revtex4-2} 

\usepackage [utf8]{inputenc}
\usepackage{graphicx}
\usepackage{dcolumn}
\usepackage{bm}

\begin{document}

\title{Density-polarity coupling in confined active polar films: \\
asters, spirals, and biphasic orientational phases}


\author{Mathieu Dedenon$^{1,2}$}
\author{Claire A. Dessalles$^1$}
\author{Pau Guillamat$^3$}
\author{Aur\'elien Roux$^1$}
\author{Karsten Kruse$^{1,2}$}
\email{karsten.kruse@unige.ch}
\author{Carles Blanch-Mercader$^4$}
\email{carles.blanch-mercader@curie.fr}
\affiliation{$^1$Department of Biochemistry, University of Geneva, 1211 Geneva, Switzerland}
\affiliation{$^2$Department of Theoretical Physics, University of Geneva, 1211 Geneva, Switzerland}
\affiliation{$^3$Institute for Bioengineering of Catalonia, Barcelona Institute of Science and Technology, Barcelona, Spain}
\affiliation{$^4$PhysicoChimie Curie, Institut Curie, PSL Research University, CNRS UMR168, Paris, France}

\date{\today}



\begin{abstract}
Topological defects in active polar fluids can organise spontaneous flows and influence macroscopic density patterns. Both of them play, for example, an important role during animal development. Yet the influence of density on active flows is poorly understood. Motivated by experiments on cell monolayers confined to discs, we study the coupling between density and polar order for a compressible active polar fluid in presence of a $+1$ topological defect. As in the experiments, we find a density-controlled spiral-to-aster transition. In addition, biphasic orientational phases emerge as a generic outcome of such coupling. Our results highlight the importance of density gradients as a potential mechanism for controlling flow and orientational patterns in biological systems.
\end{abstract}
\maketitle

Active matter is composed of individual constituents able to extract energy from their local environment to produce mechanical work \cite{marchetti_hydrodynamics_2013,shankar_topological_2022}.
This feature gives rise to collective phenomena that play an important role in many biological systems, such as the emergence of polar flocking, motility-induced phase separation or spontaneous flows \cite{marchetti_hydrodynamics_2013,shankar_topological_2022}. For instance, spontaneous flows generated by gradients of active stress have been observed in various systems, including cytoskeleton assays \cite{nedelec_self-organization_1997,schaller_polar_2010,opathalage_self-organized_2019}, or multicellular ensembles \cite{siegert_spiral_1995,kawaguchi_topological_2017,saw_topological_2017,duclos_spontaneous_2018,streichan_global_2018}.
All these systems can organize into out-of-equilibrium phases with domains featuring orientational order. This order can locally be disrupted by disclinations, often called topological defects, which are associated with rotational flow patterns~\cite{doostmohammadi_active_2018,shankar_topological_2022,leclech_physiological_2023}. 

Both, theoretical and experimental studies have demonstrated that the interplay between topological defects and active processes concentrates mechanical stress, leading to the formation of density gradients \cite{giomi_defect_2014,saw_topological_2017,kawaguchi_topological_2017,turiv_topology_2020,endresen_topological_2021,brezin_spontaneous_2022,guillamat_integer_2022,kaiyrbekov_migration_2023}.
Reciprocally, cell density variations influence orientational order \cite{duclos_perfect_2014,blanch-mercader_quantifying_2021}. Given the growing recognition of topological defects as organizing centers during morphogenesis \cite{saw_topological_2017,kawaguchi_topological_2017,maroudas-sacks_topological_2021,guillamat_integer_2022}, understanding how density gradients and orientational order interact is essential.

\begin{figure}
\includegraphics[width=0.48\textwidth]{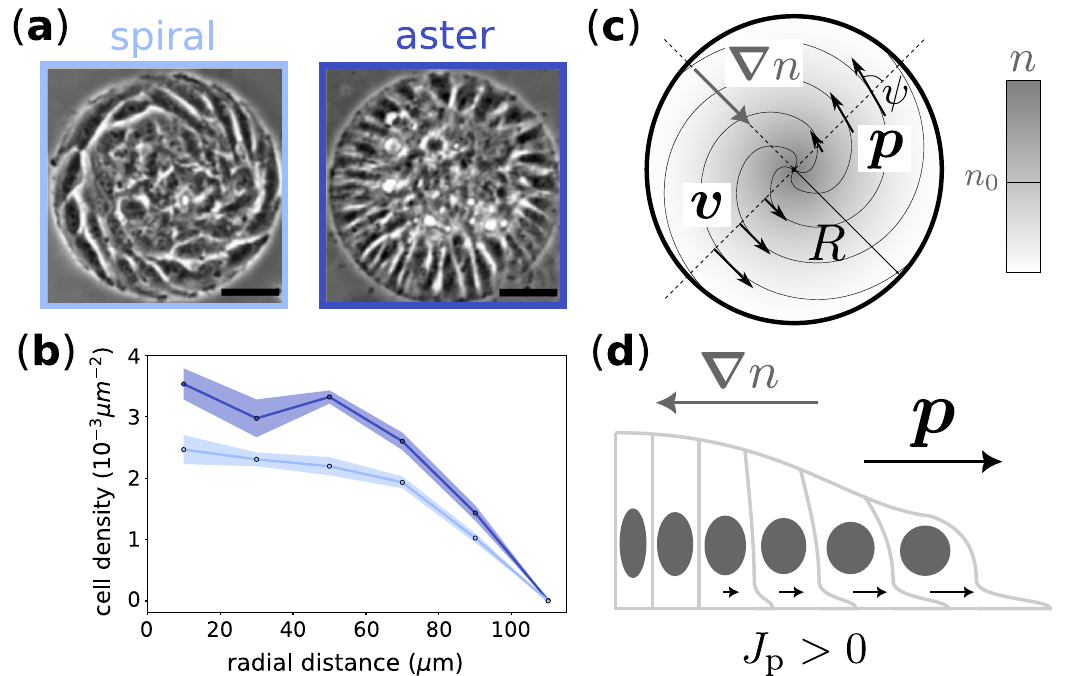}
\centering
\caption{\textit{Density-driven transition of a confined polar tissue.}
\textbf{(a)} Phase-contrast image of a confined monolayer of C2C12 myoblasts, showing a spiral (left) or an aster (right) polar state. Scale bar is $50~\mu$m. Modified from \cite{blanch-mercader_quantifying_2021}.
\textbf{(b)} Radial cell density profile for spirals and asters. Data extracted from \cite{guillamat_integer_2022}.
\textbf{(c)} Schematic representing a polar tissue confined to a disc of radius $R$, described as a 2D compressible polar fluid with velocity $\bm{v}$, polarity $\bm{p}$ with radial angle $\psi$, and density $n$.
\textbf{(d)} Schematic representing the effect of the Density-Polarity Coupling (DPC), see Eq.~\eqref{eq-fDPC}.}
\label{fig1}
\end{figure}

A density-controlled transition between different $+1$ topological defects was observed in monolayers of polarized cells confined to a disc \cite{guillamat_integer_2022}.
At low cell density, spontaneous rotational flows emerged in a spiral multicellular arrangement. Whereas for increasing cell density, a transition occurred to an aster arrangement without rotational flows, Fig.~\ref{fig1}a. Steeper cell density gradients were found for asters compared to spirals, Fig.~\ref{fig1}b.
In the hydrodynamic description of an incompressible active polar fluid, an aster-to-spiral transition arises from the competition between the active stress and orientational elasticity \cite{kruse_asters_2004}.
The transition corresponds to a spontaneous flow instability \cite{voituriez_spontaneous_2005,giomi_complex_2008,furthauer_taylorcouette_2012,shendruk_dancing_2017}, where density does not appear explicitly as a control parameter.

In this Letter, we study a coupling between density gradients and orientational order, in the case of $+1$ topological defects in confined active polar fluids, Fig.~\ref{fig1}d. In spreading cell monolayers, this Density-Polarity Coupling (DPC) expresses a tendency of cells to polarize away from high density regions \cite{streichan_spatial_2014,heinrich_size-dependent_2020,alert_physical_2020}.
First, we identify conditions for a density-controlled spiral-to-aster transition. Second, we show that biphasic orientational phases are a generic feature of compressible polar fluids. Finally, we discuss the relevance of DPC for monolayers of polarized cells.


To describe a two-dimensional compressible active polar fluid, we use active gel theory \cite{kruse_asters_2004,julicher_hydrodynamic_2018}. The system is characterised by velocity $\bm{v}(\bm{r},t)$, polarity $\bm{p}(\bm{r},t)$ and particle density $n=n_0\bar{n}(\bm{r},t)$ fields, where $n_0$ is the preferred particle density, Fig.~\ref{fig1}c.

The equilibrium physics is captured by an effective free energy $\mathcal{F}=\int_A\mathrm{d}A\,f$ with free-energy density
\begin{equation}\label{eq-ftot}
f=\frac{B}{2}\left(1-\bar{n}\right)^2+\frac{G}{2}|\bm{\nabla}\bar{n}|^2+\frac{K}{2}|\bm{\nabla}\bm{p}|^2+\frac{\chi}{2}\bm{p}^2+f_{\rm DPC}.
\end{equation}
The first two terms penalize density variations with elastic coefficients $B,G>0$. The second two terms tend to suppress polarity variations with elastic coefficients $K,\chi>0$. Thus, we favour a disordered phase in the bulk. We use the one-constant approximation \cite{gennes_physics_1993} for simplicity and leave the general case for future studies.

The last term, $f_{\rm DPC}$, accounts for the coupling between density and polarity. The lowest order term in powers of $(\bm{p},\bm{\nabla}\bar{n})$ with polar symmetry reads
\begin{equation}\label{eq-fDPC}
f_{\rm DPC}=J_{\rm p}n_0(\bm{p}\cdot\bm{\nabla})\bar{n},
\end{equation}
which is related to a density-dependent spontaneous splay term of the Frank free energy \cite{frank_liquid_1958,gennes_physics_1993,alert_physical_2020}. Previous works identified a linear instability of an ordered state associated with this coupling \cite{kung_hydrodynamics_2006,voituriez_generic_2006,adar_active-gel_2022,adar_permeation_2021,ibrahimi_deforming_2023}.
Negative (positive) values of the coupling coefficient $J_{\rm p}$ favor (anti-)alignment of polarity to density gradients, Fig.~\ref{fig1}d. From now on, we use $ J_{\rm p} n_0\equiv j_{\rm p}$ as control parameter.

The evolution of the fields $\bar{n}$, $\bm{v}$ and $\bm{p}$ is determined by the continuity equation, the polarity dynamics and the local force balance:
\begin{subequations}
\label{eq-evo}
\begin{eqnarray}
\partial_t\bar{n}&=&-\partial_{\beta}(\bar{n} v_{\beta}) \\
D_tp_{\alpha}&=&\frac{h_{\alpha}}{\gamma}-\nu\left(v_{\alpha\beta}-\frac{1}{2}v_{\gamma\gamma}\delta_{\alpha\beta}\right)p_{\beta} \\
0&=&\partial_{\beta}(\sigma_{\alpha\beta}^{\rm e}+\sigma_{\alpha\beta}^{\rm d}),
\end{eqnarray}
\end{subequations}
where $\mathbf{h}=-\delta\mathcal{F}/\delta\bm{p}$ is the molecular field, $v_{\alpha\beta}=(\partial_{\alpha}v_{\beta}+\partial_{\beta}v_{\alpha})/2$, and $\omega_{\alpha\beta}=(\partial_{\alpha}v_{\beta}-\partial_{\beta}v_{\alpha})/2$ are the symmetric and anti-symmetric parts of the velocity gradient tensor, and $D_tp_{\alpha}=\partial_tp_{\alpha}+v_{\beta}\partial_{\beta}p_{\alpha}+\omega_{\alpha\beta}p_{\beta}$ is the co-rotational derivative.
The stress is decomposed into the Ericksen and the deviatoric components that read
\begin{subequations}
\label{eq-stress}
\begin{eqnarray}
\sigma_{\alpha\beta}^{\rm e}&=&-P\delta_{\alpha\beta}-(G\partial_\beta\bar{n}+j_{\rm p}p_{\beta})\partial_\alpha\bar{n}-K\partial_\alpha p_\gamma \partial_\beta p_\gamma \\
\sigma_{\alpha\beta}^{\rm d}&=&2\eta\left(v_{\alpha\beta}-\frac{1}{2}v_{\gamma\gamma}\delta_{\alpha\beta}\right) \\ \nonumber
&+&\frac{\nu}{2}(p_{\alpha}h_{\beta}+p_{\beta}h_{\alpha}-p_{\gamma}h_{\gamma}\delta_{\alpha\beta})+\frac{1}{2}(p_{\alpha}h_{\beta}-p_{\beta}h_{\alpha}) \\ \nonumber
&-&\frac{1}{2}\zeta_0\Delta\mu p_{\gamma}p_{\gamma}\delta_{\alpha\beta}-\zeta\Delta\mu\left(p_{\alpha}p_{\beta}-\frac{1}{2}p_{\gamma}p_{\gamma}\delta_{\alpha\beta}\right)
\end{eqnarray}
\end{subequations}
with the pressure $P=\mu\bar{n}-f$, the chemical potential $\mu=\delta \mathcal{F}/\delta \bar{n}$ and $\Delta\mu$ is the chemical potential difference extracted from fuel consumption. The phenomenological parameters are the rotational viscosity $\gamma$, the flow alignment coefficient $\nu$, the shear viscosity $\eta$, and the active isotropic (anisotropic) coefficient $\zeta_0$ ($\zeta$).

As in the experimental system of Ref.~\cite{guillamat_integer_2022}, we consider an active fluid confined to a disc of radius $R$, Fig.~\ref{fig1}c. Using polar coordinates $(r,\theta)$, the polarity field is decomposed into the polar order $S$ and the tilt angle $\psi$ with respect to the radial direction, so that $\bm{p}=S\cos\psi\,\mathbf{e}_r+S\sin\psi\,\mathbf{e}_{\theta}$, where $\mathbf{e}_r$ and $\mathbf{e}_\theta$ are the unit polar vectors. In addition we assume rotational invariance, $\partial_{\theta}=0$. Because our theoretical description is achiral, without loss of generality, we restrict the range of angles to $\psi=[0,\pi]$. Using the convention that outward polarity corresponds to $\psi<\pi/2$, one can classify the different $+1$ topological defects into out-aster $\psi=0$, out-spiral $0<\psi<\pi/2$, vortex $\psi=\pi/2$, in-spiral $\pi/2<\psi<\pi$ and in-aster $\psi=\pi$.

The evolution equations for the fields $\bar{n}$, $S$, $\psi$, $v_r$ and $v_{\theta}$ are detailed in Supplementary Material (SM) \cite{supp_mater}. Motivated by the experiments in \cite{guillamat_integer_2022}, spatial boundary conditions at $r=R$ are set to $S=1$ (boundary-induced order), $\partial_r\psi=0$ (free orientation), $v_r=0$ (absence of particle flux), and $\sigma_{\theta r}=0$ (absence of shear stress). At equilibrium, the last boundary condition at $r=R$ is obtained from the minimization of the free energy \eqref{eq-ftot}, which yields $\partial_r\bar{n}=-j_{\rm p}\cos\psi/G$. We assume that this condition also holds out-of-equilibrium. 
At $r=0$, regularity of the solution imposes that $S=\partial_r\psi=\partial_r\bar{n}=v_r=v_{\theta}=0$. 

Parameters are non-dimensionalized by using disc radius $R$ as length scale, Frank constant $K$ as energy scale and rotational viscosity $\gamma$ to obtain a time scale $\gamma R^2/K$. In the following, $B=12$, $G=2$, $\eta=2$, $\nu=-1.5$ are fixed, and $\chi$, $j_{\rm p}$, $\zeta\Delta\mu$, $\zeta_0\Delta\mu$ are varied. In numerics, the initial polarity is oriented outwards (i.e. $\psi(r,t=0)<\pi/2$), and the total particle density $\int_A\mathrm{d}A\,n/A$ is set to $n_0$ to avoid any pre-stress in the uniform configuration. For more details on the numerical scheme and initial conditions, see SM \cite{supp_mater}.

\begin{figure}
\includegraphics[width=0.48\textwidth]{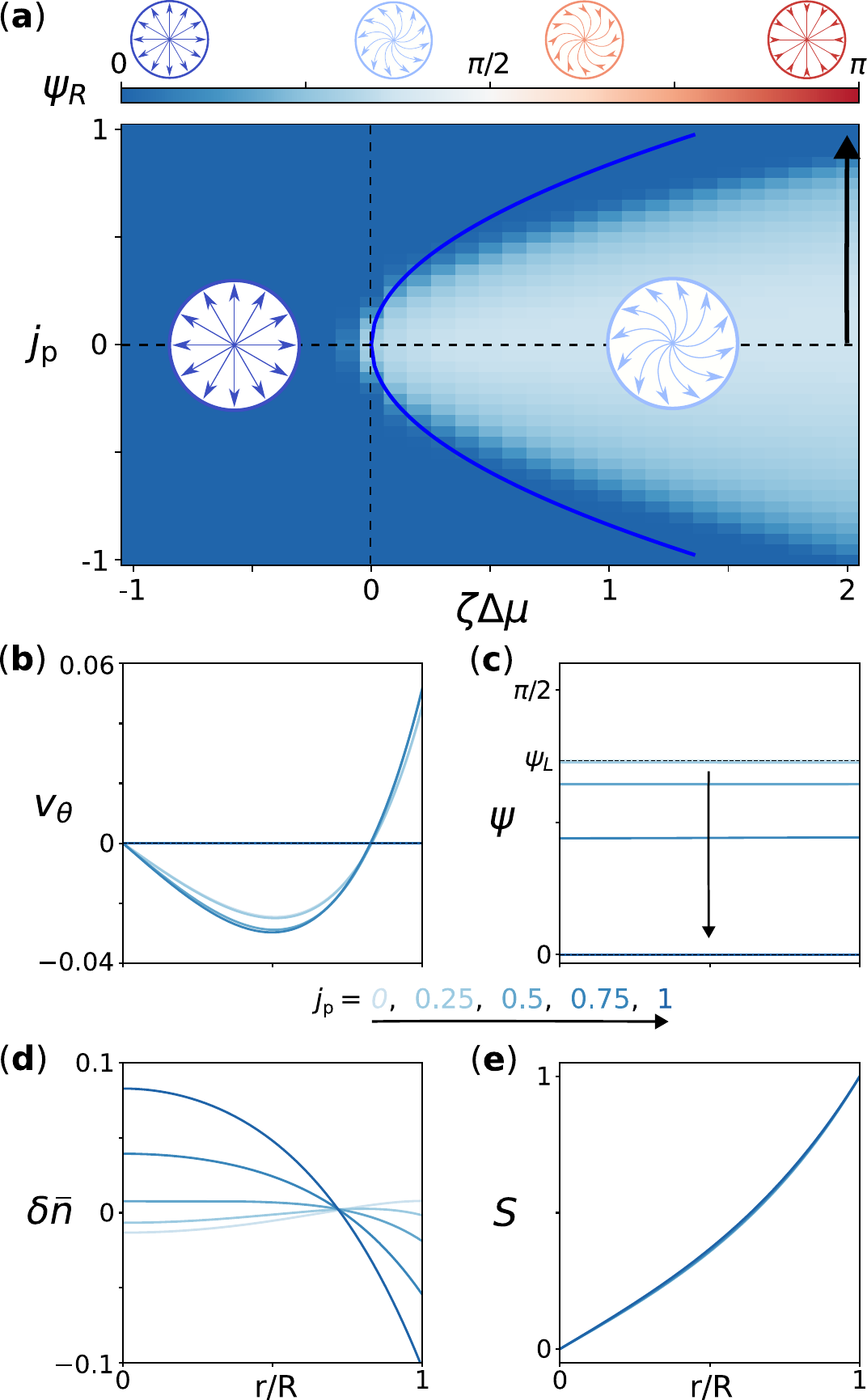}
\centering
\caption{\textit{Spiral-to-aster transition induced by DPC.}
\textbf{(a)} Density plot of the peripheral angle $\psi_R=\psi(R)$ at steady-state, as a function of anisotropic activity ($\zeta\Delta\mu$) and DPC ($j_{\rm p}$) coefficients. Blue curve: threshold $|j_{\rm p}|=|j_{\rm p}^*|$ from Eq.~\eqref{eq-threshold}.
\textbf{(b-e)} Radial profiles of azimuthal velocity $v_{\theta}(r)$ (b), angle $\psi(r)$ (c), density variation $\delta\bar{n}=\bar{n}(r)-1$ (d) and polar order $S(r)$ (e), for $\zeta\Delta\mu=2$ and $j_{\rm p}$ varies as indicated in legend (e) and black arrow (a,c). Gray line in (c): Leslie angle $\psi_L$. Parameters are $\chi=4$ and $\zeta_0\Delta\mu=0$.}
\label{fig2}
\end{figure}

First, we consider the case of vanishing activity $\zeta\Delta\mu=\zeta_0\Delta\mu=0$. In this case, the equilibrium states are in- and out- asters, Fig.~\ref{fig2}a, which have the same total energy and are selected through spontaneous symmetry breaking. The corresponding density gradients have opposite signs, see SM \cite{supp_mater}.

Next, in the case of vanishing DPC $j_{\rm p}=0$, spontaneous flows occur when $\zeta\Delta\mu>0$, Fig.~\ref{fig2}. Specifically, in- and out-asters transition to rotating spirals when anisotropic activity switches from contractile $\zeta\Delta\mu<0$ to extensile $\zeta\Delta\mu>0$, Fig.~\ref{fig2}a-c. Unlike in past works \cite{kruse_asters_2004,voituriez_spontaneous_2005}, here the instability threshold vanishes because of the absence of boundary anchoring. Spirals feature counter-rotating flows with a vanishing net torque because forces are internal, see Fig.~\ref{fig2}b. Their steady-state orientation angle $\psi(r)=\psi_L$ satisfies the relation $\nu\cos(2\psi_L)=1$ \cite{leslie_constitutive_1968}, where $\psi_L$ is the Leslie angle, see Fig.~\ref{fig2}c. Gradients of density are sustained by active processes in both spirals and asters, with their direction set by $\bar{n}'\sim-(\zeta\cos(2\psi)+\zeta_0)\Delta\mu$ for uniform $\psi$ \cite{blanch-mercader_integer_2021}, see Fig.~\ref{fig2}d for an extensile spiral.

Based on the above results, when $j_{\rm p}\neq 0$ and $\zeta\Delta\mu>0$, we expect competition between DPC, promoting radial configurations, and the active anisotropic stress driving the polarity towards the Leslie angle. Solving numerically our hydrodynamic equations \eqref{eq-evo}, a spiral-to-aster transition is found at a threshold value of $j_{\rm p}$, Fig.~\ref{fig2}a and c. As $|j_{\rm p}|$ increases near the threshold value, density gradients become steeper and the angle $\psi$ approaches zero as for the out-aster state thanks to DPC, Fig.~\ref{fig2}c,d. In contrast, the polar order parameter remains approximately independent of $j_{\rm p}$, Fig.~\ref{fig2}e. Importantly, this transition now occurs at a finite threshold of activity, Fig.~\ref{fig2}a.

To further understand this competition, we analysed the linear stability of an out-aster to perturbations in the angle $\psi$, see SM \cite{supp_mater}. Neglecting gradients of orientation Fig.~\ref{fig2}c, the linear dynamics for the angle perturbation $\delta\psi$ reduces to 
\begin{equation}\label{eq-aster}
\partial_t\delta\psi\propto\left\{j_{\rm p}\bar{n}_{\rm a}'+\frac{2\zeta\Delta\mu\gamma(1-\nu)S_{\rm a}^3}{4\eta+\gamma S_{\rm a}^2(\nu-1)^2}\right\}\delta\psi
\end{equation}
where $S_{\rm a}(r)$ and $\bar{n}_{\rm a}(r)$ are, respectively, the steady-state polar order and reduced density for an out-aster. Assuming that the instability originates from the boundary, we replace these profiles by their boundary values $S_{\rm a}=1$ and $\bar{n}_{\rm a}'=-j_{\rm p}/G$ in Eq.~\eqref{eq-aster} and obtain the analytical threshold
\begin{equation}\label{eq-threshold}
|j_{\rm p}^{*}|=\sqrt{\frac{2\zeta\Delta\mu G\gamma(1-\nu)}{4\eta+\gamma(1-\nu)^2}}.
\end{equation}
This threshold suggests that an out-aster is linearly unstable for $\zeta\Delta\mu(1-\nu)>0$ and an intermediate range of the DPC coefficient $|j_{\rm p}|<|j_{\rm p}^{*}|$. Expression~\eqref{eq-threshold} is in qualitative agreement with numerics, Fig.~\ref{fig2}a. In conclusion, DPC can suppress the spontaneous flow transition and stabilise asters in active polar fluids.

Let us reconsider the equilibrium case. There, linear stability analysis shows that DPC alone can destabilize a uniform ordered state  \cite{kung_hydrodynamics_2006,voituriez_generic_2006,adar_active-gel_2022,adar_permeation_2021}. Indeed equilibration of density fluctuations leads to an effective Frank free-energy with a renormalized splay constant $K_{\rm s}=K-j_{\rm p}^2/B$, whereas the bend constant remains unchanged $K_{\rm b}=K$, see SM \cite{supp_mater}. For $K_{\rm s}<0$, that is above the threshold value $|j_{\rm p}^{\dagger}|=\sqrt{KB}$, splay distortions are favoured. In our system, the threshold for this instability $j_{\rm p}^{\dagger}$ is modified by activity and boundary conditions. The instability is associated with a finite wavelength, which can generate biphasic orientational phases in the context of $+1$ topological defects that we analyze in the following.

\begin{figure}
\includegraphics[width=0.48\textwidth]{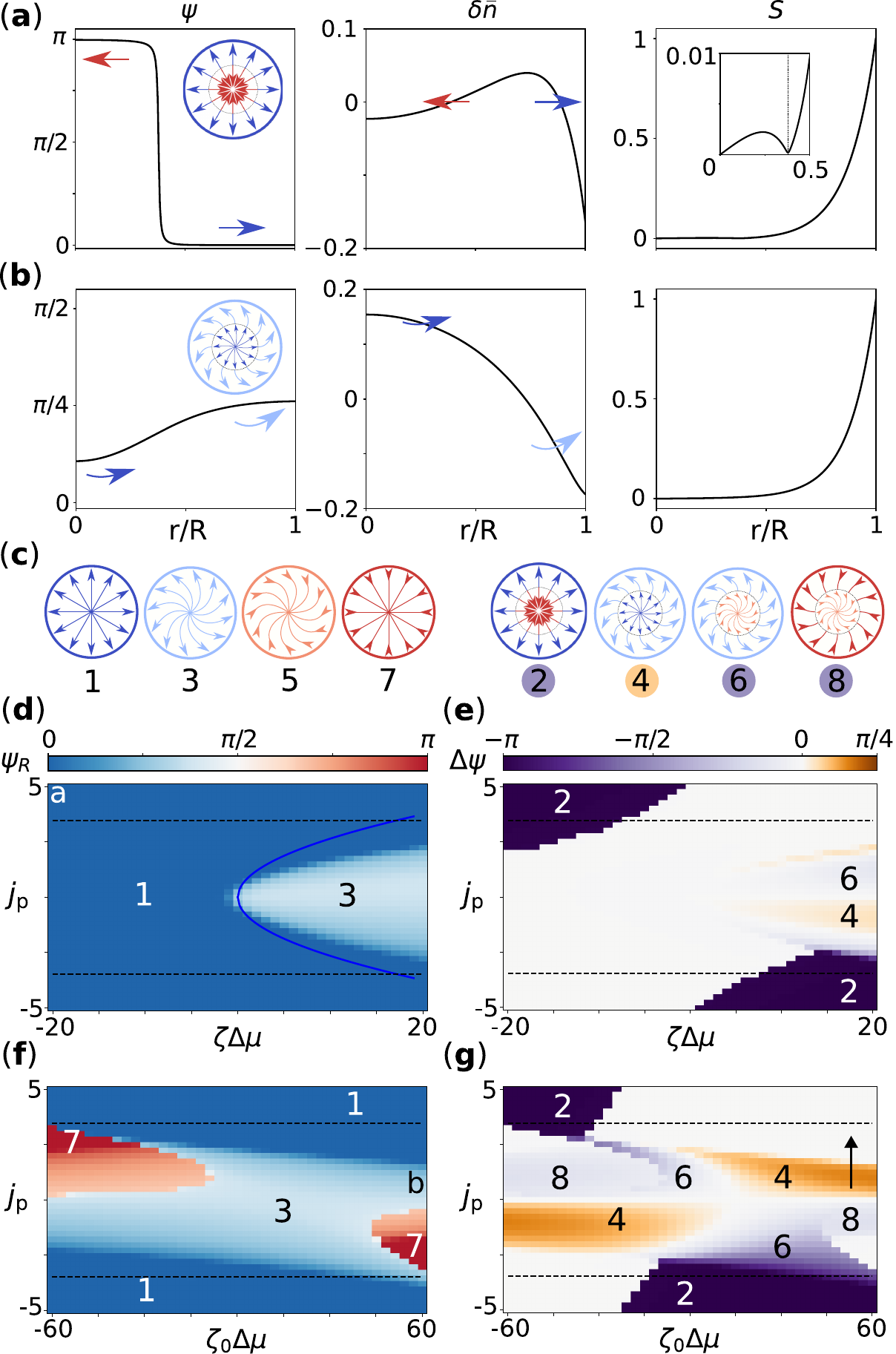}
\centering
\caption{\textit{Orientational patterns induced by DPC.}
\textbf{(a-b)} Radial profiles of angle (left), density variation (middle) and polar order (right), for biphasic asters (a) and double spirals (b). Inset: polar order near disc center.
\textbf{(c)} Schematics of orientation states.
\textbf{(d-g)} Steady-state density plots: peripheral angle $\psi_R=\psi(R)$ (d,f) and angle difference between periphery and center $\Delta\psi=\psi(R)-\psi(0)$ (e,g).
Dashed lines in (d-g): $j_{\rm p}=\pm\sqrt{KB}$. Black arrow in (g): double spiral to aster transition.
Parameters $\chi=81$, $\zeta_0\Delta\mu=0$ for (d,e), and $\zeta\Delta\mu=20$ for (f,g).}
\label{fig3}
\end{figure}

Beyond the spontaneous splay instability, biphasic asters emerge where in- and out-aster states coexist, Fig.~\ref{fig3}a. This state is characterised by a non-monotonic density profile, favouring non-uniform orientations due to DPC, Fig.~\ref{fig3}a, and a sharp interface with strong orientation gradients ($R|\boldsymbol{\nabla}\psi|\gg 1$). Because the positive bend constant $K_{\rm b}=K$ prevents large gradients of $\psi$, the polar order $S$ needs to be sufficiently small to stabilize the interface, Fig.~\ref{fig3}a, inset. This can be achieved in the disordered limit $\sqrt{K/\chi}\ll R$, such that polar order is localized at the disc periphery, see Fig.~\ref{fig3}a.

Below the spontaneous splay instability, double spirals can be found. They are characterised by a gradual gradient of orientation ($R|\boldsymbol{\nabla}\psi|\sim 1$), Fig.~\ref{fig3}b. This gradient results from a competition between active alignment and DPC, modulated by the local amplitude of polar order $S$. Indeed, if anisotropic activity dominates over DPC at the periphery ($S\sim 1$), spirals are stabilised for $\zeta\Delta\mu>0$, Fig.~\ref{fig3}b. Away from the periphery, where order is weak ($S\ll 1$), DPC always dominates, favouring out-asters for inward density gradients, Fig.~\ref{fig3}b. Contrary to Fig.~\ref{fig2}c,e where polar order remains large near the center, locally attenuating the competition between active alignment and DPC, here the disordered limit $\sqrt{K/\chi}\ll R$ results in larger orientational gradients.

These states can be characterized by the peripheral angle $\psi_R=\psi(R)$ and the angle difference between the periphery and the center $\Delta\psi=\psi(R)-\psi(0)$. Whereas Fig.~\ref{fig3}d is apparently similar to Fig.~\ref{fig2}a, the state diagram for the angle difference in Fig.~\ref{fig3}e reveals biphasic asters and double spirals with $\Delta\psi\neq0$, SM \cite{supp_mater}. The dependence on activity of the spontaneous splay threshold can be understood from the non-monotonicity of density profiles, as in Fig.~\ref{fig3}a. Whereas density gradients $\bar{n}'(R)$ are set by DPC at the periphery, in the bulk, they scale as $\bar{n}'\sim-(\zeta+\zeta_0)\Delta\mu$ when activity dominates. Therefore, biphasic asters are favoured when $j_{\rm p}>0$ and $(\zeta+\zeta_0)\Delta\mu<0$ or vice-versa, in agreement with state $2$ in Fig.~\ref{fig3}e when $\zeta_0\Delta\mu=0$, or in Fig.~\ref{fig3}g when $\zeta_0\Delta\mu\neq0$.

At low values of $|j_{\rm p}|$, double spirals can emerge, states $4$, $6$ and $8$ in Fig.~\ref{fig3}e,g. Whereas peripheral orientation remains outward when $\zeta_0\Delta\mu=0$, Fig.~\ref{fig3}e, large isotropic activity can induce inward oriented states $7$ and $8$ in Fig.~\ref{fig3}f,g. These states can no longer be understood from peripheral angle dynamics alone. They appear when anisotropic active stresses overcome DPC at the periphery, in combination with outward (inward) bulk density gradients to promote inward orientation for $j_{\rm p}>0$ ($j_{\rm p}<0$). Increasing $\zeta_0\Delta\mu$ to positive values changes the direction of bulk density gradients, and reverses the central angle from inwards to outwards through the sequence of states $8\rightarrow 6\rightarrow 4$ for $j_{\rm p}>0$, see Fig.~\ref{fig3}c,g.

In summary, a local coupling between polarity and density gradients can account for the observed transition between rotating spirals and non-flowing asters as cell density increases, Fig.~\ref{fig1}. In addition, these results provide an alternative interpretation of this transition, in terms of a transition from a double spiral to an aster, black arrow in Fig.~\ref{fig3}g. In this case, for low densities, a double spiral with aster-like orientation $\psi\simeq 0$ in the center is found, Fig.~\ref{fig3}b,g. With increasing density, this inner phase expands until it fills the entire disc and the angle becomes $\psi=0$. This double-spiral state delays the relaxation of the peripheral angle towards zero, which is consistent with experiments, see \textit{appendix}.

The spiral-to-aster transition discussed above crucially relies on the choice of the free energy term $f_{\rm DPC}$ (Eq.~\ref{eq-fDPC}). Alternatively, it can be written as $f_{\rm DPC}=-j_{\rm p}\bar{n}\boldsymbol{\nabla}\cdot\boldsymbol{p}$, which leads to the equilibrium BC $\partial_r\bar{n}(R)=0$. In SM \cite{supp_mater}, we show that the main results remain unchanged for different parameter values. One could also consider a free energy of the form $\tilde{f}_{\rm DPC}=J_{\rm p}(\bm{p}\cdot\bm{\nabla})n/n_0$ \cite{kung_hydrodynamics_2006,giomi_complex_2008,notbohm_cellular_2016}. In this case, a density-controlled spiral-to-aster transition occurs if $\zeta\Delta\mu$ decreases with density. Then, other parameters like isotropic active stress also need to depend on density to match the observed density profiles in Fig.~\ref{fig1}b. Therefore, Eq.~\ref{eq-fDPC} corresponds to a minimal extension of Ref.~\cite{blanch-mercader_quantifying_2021}.

DPC not only provides an explanation for the dynamics of polar tissues on discs, Fig.~\ref{fig1}, but also proposes a mechanism for collective states found in giant epithelial cell monolayers \cite{heinrich_size-dependent_2020}. There, a radially spreading tissue develops azimuthal flows in the central region, and density gradients become non-monotonic. In our framework, this state resembles biphasic asters except for an outward spiral orientation near the center in Ref.~\cite{heinrich_size-dependent_2020}. We expect this difference to originate from a global polar order, which is able to sustain bulk active stresses contrary to our disordered system. Validation of these hypotheses requires a precise measurement of the cell polarity field and complementary theoretical analysis.

To our knowledge, the above experimental works represent the first evidences of DPC in cellular systems. To further investigate this coupling experimentally, one could control density gradients using optogenetic tools \cite{gligorovski_multidimensional_2023} and generate specific flow or polarity patterns. Although we have focused on systems with polar symmetry, it is also interesting to consider couplings between density gradients and other types of orientational order, like nematic systems \cite{wang_patterning_2023}.





\begin{acknowledgments}
We are grateful to Jean-Fran\c cois Joanny and Ram Adar for insightful discussions, Ricard Alert for pointing out reference \cite{heinrich_size-dependent_2020}, and Ludovic Dumoulin for help on numerical methods.
The computations were performed at University of Geneva on Baobab HPC cluster.
C.A.D. acknowledges funding from the EMBO fellowship ALTF 886-2022, and
P.G. acknowledges support from the Human Frontiers of Science Program (grant number LT-000793/2018-C).
\end{acknowledgments}

\appendix*
\renewcommand{\thefigure}{A\arabic{figure}}
\setcounter{figure}{0}

\section{}
\begin{figure}[t]
\includegraphics[width=0.45\textwidth]{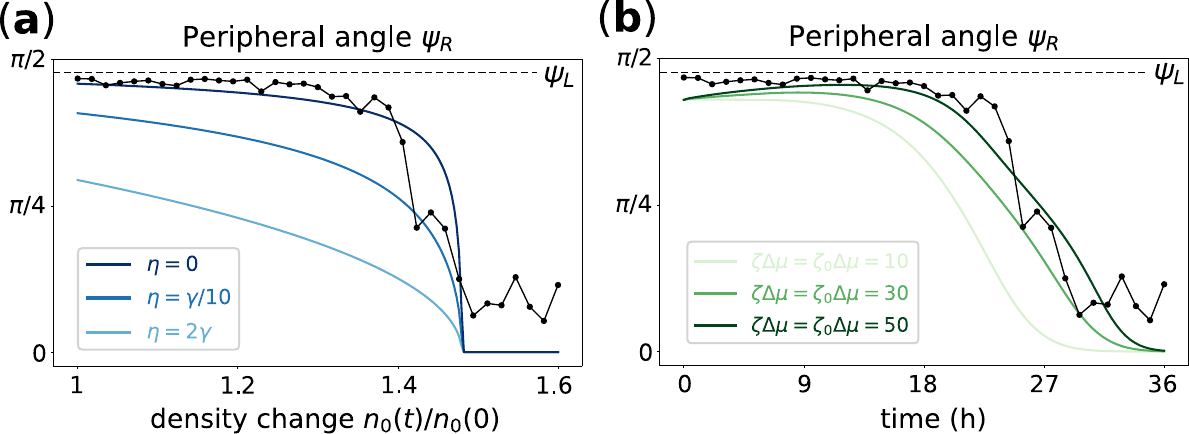}
\centering
\caption{\textit{Experimental-theoretical comparison of peripheral angle evolution.}
\textbf{(a)} Peripheral angle $\psi_R=\psi(R)$ as a function of the reduced density $\hat{n}=n_0/n_0^*$ (black: experiment, blue: Eq.~\eqref{eq:appendix1}) for different values of $\eta/\gamma$ and $\nu=-1.01$. The value of $j=j_{\rm p}/\sqrt{\zeta\Delta\mu G}$ is chosen such that the spiral-to-aster transition occurs at $\hat{n}=1.48$. Experimental densities were linearly interpolated between the initial value $n_0^*$ and the value at the transition.
\textbf{(b)} Peripheral angle $\psi_R=\psi(R)$ as a function of time (black: experiment, green: numerics) for different values of active stresses $\zeta/\Delta\mu=\zeta_0\Delta\mu$. The other parameters are $\nu=-1.01$, $\eta/\gamma=2$, $\chi=81$, $\psi(t=0)=\psi_L$ and $j_{\rm p}(t)=3.5(t/36~\text{h})$. Data extracted from \cite{guillamat_integer_2022}.}
\label{figapp}
\end{figure}

Here we compare the evolution of the peripheral angle $\psi_R$ between experiment and theory. Experiments in Ref.~\cite{guillamat_integer_2022} first show a spiral maintained over one day, followed by a rapid transition to an aster, see Fig.\ref{figapp}b. In theory, assuming uniform angle $\psi=\psi_R$ and transition controlled by boundary effects, we obtain an expression for the angle
\begin{equation}
\cos(2\psi_R)=\frac{1}{\nu}\frac{1}{1-j^2}\left[1-\frac{j^2}{2}\left(\frac{4\eta}{\gamma}+\nu^2+1\right)\right],\label{eq:appendix1}
\end{equation}
see SM \cite{supp_mater}. The critical value $j_{\rm p}^*$ at which spiral-to-aster transition occurs is given by Eq.~\eqref{eq-threshold}. Comparison between experiments and Eq.~\eqref{eq:appendix1} shows agreement for $\eta/\gamma \ll 1$, see Fig.\ref{figapp}a. Previous quantitative analysis \cite{blanch-mercader_quantifying_2021,blanch-mercader_integer_2021} suggests that $\eta/\gamma\sim 1$.

In the main text, we showed the existence of double spirals, Fig.~\ref{fig3}b. For high values of activity coefficients $\zeta\Delta\mu$, $\zeta_0\Delta\mu$ and increasing $j_{\rm p}>0$ (black arrow in Fig.~\ref{fig3}g), spiral orientation is maintained at the periphery while aster-like orientation develops in center. Compared to uniform angle states, this delays the spiral-aster transition time, see Fig.~\ref{figapp}b. Thus, double spiral itself can also quantitatively reproduce experimental data.

%

\end{document}